\documentclass[preprint,aps,12pt,showpacs,nofootinbib,tightenlines]{revtex4}
\usepackage{amsmath}
\usepackage{amssymb}
\usepackage{epsfig}
\usepackage{graphicx}
\begin{document}

\newcommand{\beq}{\begin{eqnarray}}
\newcommand{\eeq}{\end{eqnarray}}
\newcommand{\non}{\nonumber\\ }

\newcommand{\acp}{{\cal A}_{CP}}
\newcommand{\etap}{\eta^{(\prime)} }
\newcommand{\pb}{\phi_B}
\newcommand{\pp}{\phi_{\pi}}
\newcommand{\pe}{\phi_{\eta}}
\newcommand{\pepr}{\phi_{\etap}}
\newcommand{\ppp}{\phi_{\pi}^P}
\newcommand{\pep}{\phi_{\eta}^P}
\newcommand{\peprp}{\phi_{\etap}^P}
\newcommand{\ppt}{\phi_{\pi}^t}
\newcommand{\pet}{\phi_{\eta}^t}
\newcommand{\peprt}{\phi_{\etap}^t}
\newcommand{\fb}{f_B }
\newcommand{\fpi}{f_{\pi} }
\newcommand{\feta}{f_{\eta} }
\newcommand{\fetap}{f_{\etap} }
\newcommand{\rpi}{r_{\pi} }
\newcommand{\re}{r_{\eta} }
\newcommand{\rep}{r_{\etap} }
\newcommand{\mb}{m_B }
\newcommand{\mop}{m_{0\pi} }
\newcommand{\moe}{m_{0\eta} }
\newcommand{\moep}{m_{0\etap} }

\newcommand{\psl}{ P \hspace{-2.4truemm}/ }
\newcommand{\nsl}{ n \hspace{-2.2truemm}/ }
\newcommand{\vsl}{ v \hspace{-2.2truemm}/ }
\newcommand{\epsl}{\epsilon \hspace{-1.8truemm}/\,  }

\def \epjc{ Eur. Phys. J. C }
\def \jpg{  J. Phys. G }
\def \npb{  Nucl. Phys. B }
\def \plb{  Phys. Lett. B }
\def \pr{  Phys. Rep. }
\def \prd{  Phys. Rev. D }
\def \prl{  Phys. Rev. Lett.  }
\def \zpc{  Z. Phys. C  }
\def \jhep{ J. High Energy Phys.  }

\title{Branching Ratio and CP Asymmetry of $B_s \to
\pi^0 \eta^{(\prime)}$ Decays in the perturbative QCD Approach}
\author{Zhenjun Xiao$^{a}$\footnote{Electronic address: xiaozhenjun@njnu.edu.cn},
Xin Liu$^{b,a}$, Huisheng Wang$^{a}$}
\affiliation{{\it $a$ Department of
Physics and Institute of Theoretical Physics, Nanjing Normal
University, Nanjing, Jiangsu 210097, P.R.China} \\
{\it $b$ Department of Physics, Zhejiang Ocean University, Zhoushan, Zhejiang
316004, P.R.China}  } 
\date{\today}
\begin{abstract}
We calculate the branching ratios and CP-violating asymmetries for
$B_s^0 \to \pi^0 \eta^{(\prime)}$ decays in the perturbative QCD
(pQCD) factorization approach here. We not only calculate the
usual factorizable contributions, but also evaluate the
non-factorizable and annihilation type contributions.
The pQCD predictions for the CP-averaged branching ratios are
$BR(B_s^0 \to \pi^0 \eta) \approx 0.86 \times 10^{-7}$ and
$BR(B_s^0 \to \pi^0 \eta') \approx 1.86 \times 10^{-7}$.
The pQCD predictions for the CP-violating asymmetries are
$A_{CP}^{dir}(\pi^0 \eta)\sim -4.5 \% $,
$A_{CP}^{dir}(\pi^0 \eta')\sim -9.1\% $,
$A_{CP}^{mix}(\pi^0 \eta)\sim -0.2\%$, and
$A_{CP}^{mix}(\pi^0 \eta')\sim 27.0\%$ but with large errors.
The above pQCD predictions can be tested in the near future LHC-b experiments at
CERN and the BTeV experiments at Fermilab.
\end{abstract}

\vspace{1cm}


\maketitle

\section{Introduction}

The experimental measurements and theoretical studies of the two
body charmless hadronic B meson decays play an important role in
the precision test of the standard mode (SM) and in searching for
the new physics beyond the SM \cite{cpv}. For these charmless B
meson decays, the dominant theoretical error comes from the large
uncertainty in evaluating the hadronic matrix elements $\langle
M_1 M_2|O_i|B\rangle$ where $M_1$ and $M_2$ are light final state
mesons. The QCD factorization (QCDF) approach \cite{bbns99} and
the perturbative QCD (pQCD) factorization approach
\cite{cl97,li2003} are the popular methods being used to calculate the hadronic matrix elements.

When the LHC experiment is approaching, the studies about the decays of $B_s$ meson
draw much more attentions then ever before.
At present, some two-body charmless hadronic $B_s$ meson decays have been calculated, for example, in
both the QCDF  approach \cite{bn03b} and/or in the pQCD approach \cite{chen01}.
In this paper, we would like to calculate the branching ratios and
CP asymmetries for $B_s \to \pi^0 \eta^{(\prime)}$ decays
by employing the low energy effective Hamiltonian \cite{buras96}
and the pQCD factorization approach.
Besides the usual factorizable contributions, we here are able to evaluate the non-factorizable
and the annihilation contributions to these decays.

Theoretically, the two $B_s \to \pi^0 \eta^{(\prime)}$ decays have been studied
in the naive and generalized factorization approach \cite{du93,zhang01} or in the
QCD factorization approach \cite{sunbs03}.
On the experimental side, only the poor upper limits for the branching ratios are available
now \cite{pdg04}
\beq
BR(B_s^0 \to \pi^0 \etap)< 1.0 \times 10^{-3}, \label{eq:ulimits}
\eeq
Of course, this situation will be improved rapidly when LHCb starts to run at the year of 2007.

For $B_s \to \pi^0 \eta^{(\prime)}$ decays,
the light final state mesons are moving very fast in the rest frame of
$B_s$ meson. In this case, the short distance hard process dominates
the decay amplitude, while the soft final state interaction is not important for such decays,
since there is not enough time for light mesons to exchange soft gluons.
Therefore, it makes the pQCD reliable in calculating the $B_s \to
\pi^0 \eta^{(\prime)}$ decays. With the Sudakov resummation, we
can include the leading double logarithms for all loop diagrams,
in association with the soft contribution.

This paper is organized as follows. In Sec.~\ref{sec:f-work}, we
calculate analytically the related Feynman diagrams and present
the various decay amplitudes for the studied decay modes. In
Sec.~\ref{sec:n-d}, we show the numerical results for the
CP-averaged branching ratios and CP asymmetries of $B_s \to \pi^0 \eta^{(')}$
decays and compare them with the measured values or the
theoretical predictions in QCDF approach. The summary and some discussions are included
in the final section.

\section{Perturbative Calculations}\label{sec:f-work}

For $B_s \to \pi^0 \etap$ decays, the related weak effective
Hamiltonian $H_{eff}$ can be written as \cite{buras96}
\beq
\label{eq:heff} {\cal H}_{eff} = \frac{G_{F}} {\sqrt{2}} \, \left[
V_{ub} V_{us}^* \left (C_1(\mu) O_1^u(\mu) + C_2(\mu) O_2^u(\mu)
\right) - V_{tb} V_{ts}^* \, \sum_{i=3}^{10} C_{i}(\mu) \,O_i(\mu)
\right] \; .
\eeq
The explicit expressions of the operators $O_i$ can be found for example in Refs.\cite{liu05,wang05}.

In the pQCD approach, the decay amplitude is conceptually written as the convolution,
\beq
{\cal A}(B_s \to M_1 M_2)\sim \int\!\!
d^4k_1 d^4k_2 d^4k_3\ \mathrm{Tr} \left [ C(t) \Phi_{B_s}(k_1)
\Phi_{M_1}(k_2) \Phi_{M_2}(k_3) H(k_1,k_2,k_3, t) \right ],
\label{eq:con1}
\eeq
where $k_i$'s are momenta of light quarks
included in each mesons, and $\mathrm{Tr}$ denotes the trace over
Dirac and color indices. $C(t)$ is the Wilson coefficient which
results from the radiative corrections at short distance.
The function $H(k_1,k_2,k_3,t)$ describes the four quark operator and the
spectator quark connected by
 a hard gluon whose $q^2$ is in the order
of $\bar{\Lambda} M_{B_s}$, and includes the
$\mathcal{O}(\sqrt{\bar{\Lambda} M_{B_s}})$ hard dynamics.
Therefore, this hard part $H$ can be perturbatively calculated.
The function $\Phi_M$ is the wave function which describes
hadronization of the quark and anti-quark to the meson $M$. While
the function $H$ depends on the processes considered, the wave
function $\Phi_M$ is independent of the specific processes. Using
the wave functions determined from other well measured processes,
one can make quantitative predictions here.

Since the b quark is rather heavy we consider the $B_s$ meson at
rest for simplicity. It is convenient to use light-cone coordinate
$(p^+, p^-, {\bf p}_T)$ to describe the meson's momenta,
\beq
p^\pm = \frac{1}{\sqrt{2}} (p^0 \pm p^3), \quad and \quad {\bf
p}_T = (p^1, p^2).
\eeq
Using these coordinates the $B_s$ meson
and the two final state meson momenta can be written as
\beq P_1 =\frac{M_{B_s}}{\sqrt{2}} (1,1,{\bf 0}_T), \quad P_2 =
\frac{M_{B_s}}{\sqrt{2}}(1,0,{\bf 0}_T), \quad P_3 =
\frac{M_{B_s}}{\sqrt{2}} (0,1,{\bf 0}_T),
\eeq
respectively, here the light meson masses have been neglected. Putting the light
(anti-) quark momenta in $B_s$, $\pi^0$ and $\etap$ mesons as
$k_1$, $k_2$, and $k_3$, respectively, we can choose
\beq
k_1 =(x_1 P_1^+,0,{\bf k}_{1T}), \quad k_2 = (x_2 P_2^+,0,{\bf
k}_{2T}), \quad k_3 = (0, x_3 P_3^-,{\bf k}_{3T}).
\eeq
Then, the integration over $k_1^-$, $k_2^-$, and $k_3^+$ in
eq.(\ref{eq:con1}) will lead to
\beq
{\cal A}(B_s \to \pi^0 \etap)&\sim &\int\!\! d x_1 d x_2 d x_3 b_1 d b_1 b_2 d b_2 b_3 d b_3
\non && \cdot \mathrm{Tr} \left [ C(t) \Phi_{B_s}(x_1,b_1)
\Phi_{\pi^0}(x_2,b_2) \Phi_{\etap}(x_3, b_3) H(x_i, b_i, t)
S_t(x_i)\, e^{-S(t)} \right ], \label{eq:a2}
\eeq
where $b_i$ is the conjugate space coordinate of $k_{iT}$, and $t$ is the largest
energy scale in function $H(x_i,b_i,t)$.
The large double logarithms ($\ln^2 x_i$) on the longitudinal
direction are summed by the threshold resummation \cite{li02}, and
they lead to $S_t(x_i)$ which smears the end-point singularities
on $x_i$. The last term, $e^{-S(t)}$, is the Sudakov form factor
which suppresses the soft dynamics effectively \cite{soft}.
In numerical calculations, we use $\alpha_s=4\pi/[\beta_1
\ln(t^2/{\Lambda_{QCD}^{(5)}}^2)]$ which is the leading order
expression with $\Lambda_{QCD}^{(5)}=193$MeV, derived from
$\Lambda_{QCD}^{(4)}=250$MeV. Here $\beta_1=(33-2n_f)/12$, with
the appropriate number of active quarks $n_f$.

Similar to $B \to \rho \etap$ and $B \to \pi \etap$ decays, there are 8 type
diagrams contributing to the $B_s \to \pi^0 \eta^{(\prime)}$
decays, as illustrated in Figure 1. We first calculate the usual
factorizable diagrams (a) and (b). Operators $O_1$, $O_2$, $O_3$,
$O_4$, $O_9$, and $O_{10}$ are $(V-A)(V-A)$ currents, the sum of
their amplitudes is given as
\beq
F_{e\eta}&=& 4 \sqrt{2}\pi G_F
C_F f_\pi m_{B_s}^4\int_0^1 d x_{1} dx_{3}\, \int_{0}^{\infty} b_1
db_1 b_3 db_3\, \phi_{B_s}(x_1,b_1) \non & &
\times \left\{
\left[(1+x_3) \phi_{\eta}^A(x_3, b_3) +(1-2x_3) r_{\eta}^s
(\phi_{\eta}^P(x_3,b_3) +\phi_{\eta}^T(x_3,b_3))\right] \right.
\non && \left.\quad  \cdot \alpha_s(t_e^1)\,
h_e(x_1,x_3,b_1,b_3)\exp[-S_{ab}(t_e^1)] \right.\non && \left. +2
r_{\eta}^s \phi_{\eta}^P (x_3, b_3)
\alpha_s(t_e^2)h_e(x_3,x_1,b_3,b_1)\exp[-S_{ab}(t_e^2)] \right\},
\label{eq:ab}
 \eeq
where $r_{\eta}^s=m_0^{\eta_{s\bar s}}/m_{B_s}$; $C_F=4/3$ is a
color factor.The function $h_e$, the scales $t_e^i$ and the
Sudakov factors $S_{ab}$ are displayed in Appendix \ref{sec:aa}.
In the above equation, we do not include the Wilson coefficients
of the corresponding operators, which are process dependent. They
will be shown later  for different decay channels.

\begin{figure}[t,b]
\vspace{-2cm} \centerline{\epsfxsize=21 cm \epsffile{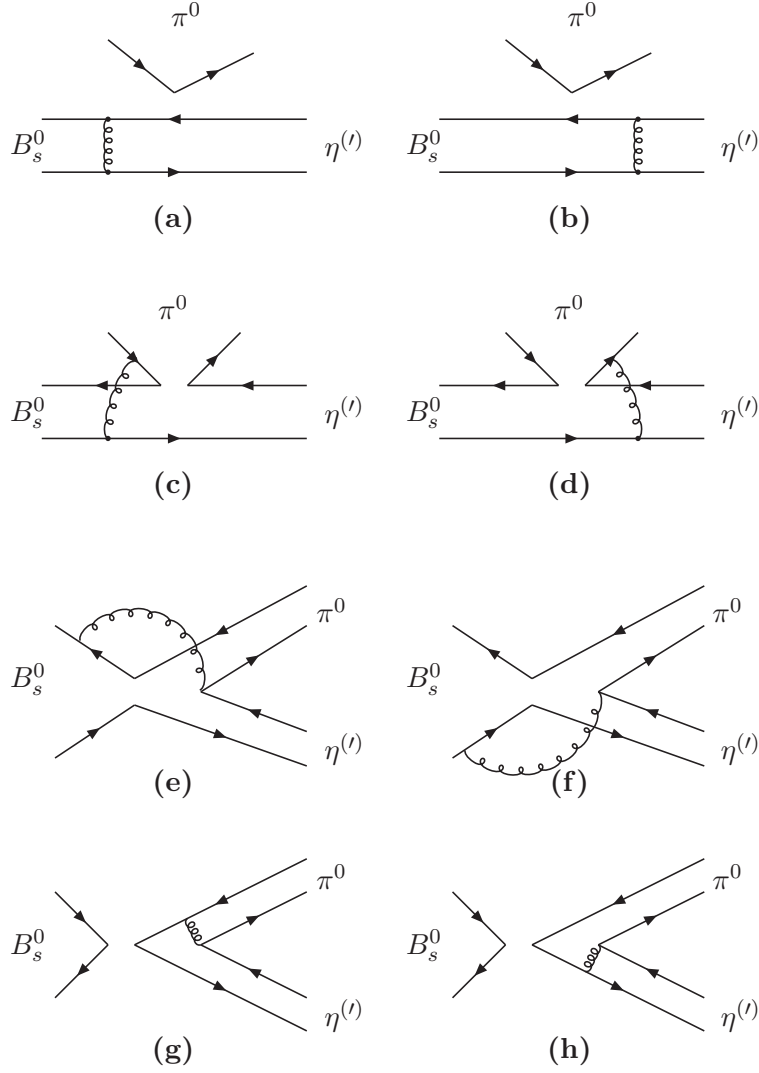}}
\vspace{-14cm} \caption{ Diagrams contributing to the $B_s^0 \to
\pi^0\eta^{(\prime)}$
 decays (diagram (a) and (b) contribute to the $B_s \to \etap$ form
 factor $F_{0,1}^{B_s^0 \to \etap}$.}
 \label{fig:fig1}
\end{figure}

The form factors of $B_s$ to $\etap$ decay, $F_{0,1}^{B_s \to
\etap}(0)$, can thus be extracted from Eq.~(\ref{eq:ab}), that is
\beq
F_{0,1}^{B_s \to \etap}(q^2=0)=\frac{F_{e\etap}}{\sqrt{2}G_F f_\pi M_{B_s}^2} .
\label{eq:f01}
\eeq

The operators $O_5$, $O_6$, $O_7$, and $O_8$ have a structure of
$(V-A)(V+A)$. In some decay channels, some of these operators
contribute to the decay amplitude in a factorizable way. Since only
the axial-vector part of $(V+A)$ current contribute to the
pseudo-scaler meson production,
$ \langle \pi |V-A|B\rangle \langle \eta |V+A | 0 \rangle = -\langle
\pi |V-A |B  \rangle \langle \eta |V-A|0 \rangle,$ that is
 \beq
 F_{e\eta}^{P1}=-F_{e\eta}\; .
 \eeq

For the non-factorizable diagrams 1(c) and 1(d), all three meson
wave functions are involved. The integration of $b_3$ can be
performed using $\delta$ function $\delta(b_3-b_2)$, leaving only
integration of $b_1$ and $b_2$.
For the $(V-A)(V-A)$ operators, the result is
\beq
 M_{e\eta}&=& \frac{16}{\sqrt{3}}\pi G_F C_F m_{B_s}^4
\int_{0}^{1}d x_{1}d x_{2}\,d x_{3}\,\int_{0}^{\infty} b_1d b_1
b_2d b_2\, \phi_{B_s}(x_1,b_1)  \non
 & &\times \phi_{\pi}^A(x_2,b_2) x_3
 \left[\phi_{\eta}^A (x_3,b_2)-2 r_{\eta}^s \phi_{\eta}^T(x_3,b_2)  \right
] \non
 & & \cdot \alpha_s(t_f) h_f(x_1,x_2,x_3,b_1,b_2)\exp[-S_{cd}(t_f)]   \; .
\eeq

$M_{e\eta}^{P}$ is for the $(S-P)(S+P)$ type
operators,which are from Fierz transformation for $(V-A)(V+A)$
operators:
\beq
M_{e\eta}^{P} &=& - M_{e\eta} .
\eeq

For the non-factorizable annihilation diagrams 1(e) and 1(f), again
all three wave functions are involved. Here we have two kinds of
contributions. $M_{a\eta}$ and $M_{a\eta}^{P}$ describe the
contributions from the $(V-A)(V-A)$ and $(S-P)(S+P)$ type
operators, respectively,
 \beq
 M_{a\eta}&=& -
\frac{16}{\sqrt{3}}\pi G_F C_F m_{B_s}^4\int_{0}^{1}d x_{1}d
x_{2}\,d x_{3}\,\int_{0}^{\infty} b_1d b_1 b_2d b_2\,
\phi_{B_s}(x_1,b_1)\non && \times \left \{\left \{ x_3
\phi_{\eta}^A(x_3,b_2) \phi_{\pi}^A(x_2,b_2) + r_\pi
r_{\eta}^{u,d} \left [ x_2
\left(\phi_{\pi}^P(x_2,b_2)-\phi_{\pi}^T(x_2,b_2)\right)\cdot
\right.\right.\right. \non
 && \left. \left.\left.
\left(\phi_{\eta}^P(x_3,b_2)-\phi_{\eta}^T(x_3,b_2)\right)+ x_3
\left(\phi_{\pi}^P(x_2,b_2)+\phi_{\pi}^T(x_2,b_2)\right)\cdot\right.\right.\right.
\non && \left.\left.\left.
\left(\phi_{\eta}^P(x_3,b_2)+\phi_{\eta}^T(x_3,b_2)\right)
 \right ] \right \}
\alpha_s(t_f^1)h_f^1(x_1,x_2,x_3,b_1,b_2)\exp[-S_{ef}(t_f^1)]
\right.  \non && \left.
  -\left \{ x_2 \phi_{\pi}^A(x_3, b_2) \phi_{\eta}^A(x_2,b_2)
 +r_\pi r_{\eta}^{u,d} \left [\left( (x_2+x_3+2) \phi_{\eta}^P(x_2,b_2)
 +(x_2-x_3) \right.\right.\right.\right.
 \non
  && \left.\left.\left.\left.\times\phi_{\eta}^T(x_2,b_2)\right)  \phi_{\pi}^P(x_3,b_2)
  + \left((x_2-x_3)\phi_{\eta}^P(x_3,b_2) +(x_2+x_3-2)\right.\right.\right.\right.
  \non
  &&\left.\left.\left.\left.\phi_{\eta}^T(x_3,b_2)\right
  )\phi_{\pi}^T(x_2,b_2)\right]
\right\}
\alpha_s(t_f^2)h_f^2(x_1,x_2,x_3,b_1,b_2)\exp[-S_{ef}(t_f^2) ]
 \right \}\; ,
 \eeq
 \beq
 M_{a\eta}^{P}&=& \frac{16}{\sqrt{3}}\pi G_F C_F m_{B_s}^4\;
 \int_{0}^{1}d x_{1}d x_{2}\,d x_{3}\,\int_{0}^{\infty} b_1d
b_1 b_2d b_2\, \phi_{B_s}(x_1,b_1)\non &&
 \times \left \{
 \left \{ x_2 \phi_{\eta}^A(x_3,b_2)
\phi_{\pi}^A(x_2,b_2) + r_\pi r_{\eta}^{u,d} \left [ x_3
 \left(\phi_{\pi}^P(x_2,b_2)-\phi_{\pi}^T(x_2,b_2)\right)\cdot
\right.\right.\right. \non
 &&\left.\left.\left.\left(\phi_{\eta}^P(x_3,b_2)-\phi_{\eta}^T(x_3,b_2)\right)+ x_2
\left(\phi_{\pi}^P(x_2,b_2)+\phi_{\pi}^T(x_2,b_2)\right)\cdot\right.\right.\right.
\non
 &&\left.\left.\left. \left(\phi_{\eta}^P(x_3,b_2)+\phi_{\eta}^T(x_3,b_2)\right)
 \right ] \right \}\alpha_s(t_f^1) h_f^1(x_1,x_2,x_3,b_1,b_2)\exp[-S_{ef}(t_f^1)]
 \right. \non
 && \left. -\left \{ x_2 \phi_{\pi}^A(x_3, b_2) \phi_{\eta}^A(x_2,b_2)
 +r_\pi r_{\eta}^{u,d} \left [\left( (x_2+x_3+2) \phi_{\eta}^P(x_2,b_2)
 +(x_3-x_2)\right.\right.\right.\right.
 \non
  && \left.\left.\left.\left.\times
  \phi_{\eta}^T(x_2,b_2)\right)  \phi_{\pi}^P(x_3,b_2)
  + \left((x_3-x_2)\phi_{\eta}^P(x_3,b_2)
  +(x_2+x_3-2)\right.\right.\right.\right.
  \non
  && \left.\left.\left.\left.
  \phi_{\eta}^T(x_3,b_2)\right
  )\phi_{\pi}^T(x_2,b_2)\right]
\right\} \alpha_s(t_f^2)
 h_f^2(x_1,x_2,x_3,b_1,b_2)\exp[-S_{ef}(t_f^2)] \right \}\; .
\eeq
where $r_{\eta}^{u,d}= m_0^{\eta_{u\bar u,d\bar d}}/m_{B_s}$.

The factorizable annihilation diagrams 1(g) and 1(h) involve only
$\pi^0$ and $\etap$ wave functions. There are also two kinds of
decay amplitudes for these two diagrams. $F_{a\eta}$ is for
$(V-A)(V-A)$ type operators, $F_{a\eta}^{P}$ is for $(V-A)(V+A)$
type operators:
\beq
F_{a\eta}^{P}=F_{a\eta}&=& 4 \sqrt{2}\pi G_F
C_F f_{B_s} m_{B_s}^4\int_{0}^{1}dx_{2}\,d x_{3}\,
\int_{0}^{\infty} b_2d b_2b_3d b_3 \, \left\{ \left[x_3
\phi_{\eta}^A(x_3,b_3) \phi_{\pi}^A(x_2,b_2)\right.\right.\non
&&\left.\left.+2 r_\pi
r_{\eta}^{u,d}((x_3+1)\phi_{\eta}^P(x_3,b_3)+(x_3-1)
\phi_{\eta}^T(x_3,b_3)) \phi_{\pi}^P(x_2,b_2)\right] \right. \non
&& \left. \quad \cdot \alpha_s(t_e^3)
h_a(x_2,x_3,b_2,b_3)\exp[-S_{gh}(t_e^3)] \right. \non && \left.
-\left[ x_2 \phi_{\eta}^A(x_3,b_3) \phi_{\pi}^A(x_2,b_2)
\right.\right.\non && \left. \left. \quad +2 r_\pi r_{\eta}^{u,d}
\phi_{\eta}^P(x_3,b_3)((x_2+1)\phi_{\pi}^P(x_2,b_2)+(x_2-1)
\phi_{\pi}^T(x_2,b_2) )\right] \right. \non &&\left. \quad \cdot
\alpha_s(t_e^4)
 h_a(x_3,x_2,b_3,b_2)\exp[-S_{gh}(t_e^4)]\right \}\; .\label{eq:mapip2}
\eeq

If we exchange the $\pi$ and $\etap$ in Fig.~1, the corresponding
expressions of amplitudes for new diagrams will be similar with
those as given in Eqs.(\ref{eq:ab}-\ref{eq:mapip2}). The
expressions of amplitudes for new diagrams can be obtained by the
replacements,
\beq
\phi_{\pi}^A \leftrightarrow \phi_{\eta}^A, \quad \phi_{\pi}^P \leftrightarrow \phi_{\eta}^P , \quad
\phi_{\pi}^T \leftrightarrow \phi_{\eta}^T, \quad r_\pi
\leftrightarrow r_{\eta}^{u,d}.
\eeq
For example, we find that
\beq
F_{a\pi}= -F_{a\eta^{(\prime)}}, \quad F_{a\pi}^{P}= -F_{a\eta^{(\prime)}}^{P}\;.
\eeq

Now we are able to calculate perturbatively the form factors
$F_0^{B_s\to \eta^{(')}}(0)$ and the decay amplitudes for the
Feynman diagrams after the integration over $x_i$ and $b_i$.Since
we here calculated the form factors and amplitudes at the leading
order ( one order of $\alpha_s(t)$), the radiative corrections at
the next order would emerge in terms of $\alpha_s(t) \ln(m/t)$,
where $m'$s denote some scales, like $m_{B_s}, 1/b_i, \ldots$, in
the hard part $H(t)$. We select the largest energy scale among
$m'$s appearing in each diagram as the hard scale $t'$s for the
purpose of at least  killing the large logarithmic corrections
partially,
\beq t_{e}^1 &=& a_t \cdot {\rm max}(\sqrt{x_3}
m_{B_s},1/b_1,1/b_3)\;,\label{t1}\non
t_{e}^2 &=& a_t \cdot {\rm
max}(\sqrt{x_1}m_{B_s},1/b_1,1/b_3)\;,\non
t_{e}^3 &=& a_t \cdot
{\rm max}(\sqrt{x_3}m_{B_s},1/b_2,1/b_3)\;,\non
t_{e}^4 &=& a_t \cdot {\rm max}(\sqrt{x_2}m_{B_s},1/b_2,1/b_3)\;,\non
t_{f} &=& a_t \cdot {\rm max}(\sqrt{x_1 x_3}m_{B_s}, \sqrt{x_2 x_3}
m_{B_s},1/b_1,1/b_2)\;,\non
t_{f}^1 &=& a_t \cdot {\rm max}(\sqrt{x_2 x_3} m_{B_s},1/b_1,1/b_2)\;,\non
 t_{f}^2 &=& a_t \cdot {\rm max}(\sqrt{x_1+x_2+x_3-x_1 x_3-x_2 x_3}m_{B_s}, \sqrt{x_2 x_3} m_{B_s},1/b_1,1/b_2)\;,
\label{tf}
\eeq
where the constant $a_t=1.0\pm 0.2$ is introduced in order to estimate the scale dependence of the
theoretical predictions for the observables.

In Ref.\cite{liu05,wang05}, a brief discussion about the
$\eta-\eta^\prime$ mixing and the gluonic component of the
$\eta^\prime$ meson have been given. Here we don't show it again.


Combining the contributions from different diagrams, the total
decay amplitude for $B_s^0 \to \pi^0 \eta$ can be written as
\beq
\sqrt{6}{\cal M}(\pi^0 \eta) &=& F_{e\eta} \left \{ \xi_u
\left(C_1+ \frac{1}{3}C_2\right) - \xi_t \left
(-\frac{3}{2}C_7-\frac{1}{2}C_8+\frac{3}{2}C_9
+\frac{1}{2}C_{10}\right)\right\}  F_2(\theta_p)\non
&&  +
M_{e\eta}\left \{ \xi_u C_2 -\xi_t \left(-\frac{3}{2}C_8
+\frac{3}{2}C_{10}\right) \right\} F_2(\theta_p) + \biggl
(M_{a\eta}+M_{a\pi}\biggr)\non
&& \cdot \left\{ \xi_u C_2-\xi_t
\frac{3}{2}C_{10}\right\} F_1(\theta_p) -\xi_t \biggl(
M_{a\eta}^{P}+M_{a\pi}^{P}\biggr)
 \frac{3}{2} C_8 F_1(\theta_p)\;  . \label{eq:m2}
\eeq

The decay amplitudes for $B_s^0 \to \pi^0 \eta'$ can be obtained
easily from Eqs.(\ref{eq:m2}) by the following replacements
\beq
F_1(\theta_p) &\longrightarrow & F'_1(\theta_p) = \cos{\theta_p} +
\frac{\sin{\theta_p}}{\sqrt{2}}, \non F_2(\theta_p)
&\longrightarrow & F'_2(\theta_p) = \cos{\theta_p} - \sqrt{2}
\sin{\theta_p}.
\eeq
Note that the possible gluonic component of $\eta'$ meson has been neglected here.

\section{Numerical results and Discussions}\label{sec:n-d}

\subsection{Input parameters and wave functions}

We use the following input parameters in the numerical
calculations
\beq
\Lambda_{\overline{\mathrm{MS}}}^{(f=4)} &=& 250 {\rm MeV}, \quad
f_\pi = 130 {\rm MeV}, \quad f_{B_s} = 230 {\rm MeV}, \non
m_0^{\eta_{d\bar{d}}}&=& 1.4 {\rm GeV},\quad
m_0^{\eta_{s\bar{s}}} = 1.95 {\rm GeV}, \quad f_K = 160  {\rm MeV}, \non
M_{B_s} &=& 5.37 {\rm GeV}, \quad M_W = 80.41{\rm GeV}.
\label{para}
\eeq
For the CKM matrix elements, here we adopt the Wolfenstein
parametrization for the CKM matrix, and take $\lambda=0.22, A=0.853, \rho=0.20$ and $\eta=0.33$
\cite{pdg04}.

For the $B_s$ meson wave function, we adopt the model
\beq
\phi_{B_s}(x,b) &=& N_{B_s} x^2(1-x)^2 \mathrm{exp} \left
 [ -\frac{M_{B_s}^2\ x^2}{2 \omega_{b_s}^2} -\frac{1}{2} (\omega_{b_s} b)^2\right],
 \label{phib}
\eeq
where $\omega_{b_s}$ is a free parameter and we take
$\omega_{b_s}=0.50\pm 0.05$ GeV in numerical calculations, and
$N_{B_s}=63.7$ is the normalization factor for
$\omega_{b_s}=0.50$.

For the light meson wave function, we neglect the $b$ dependant
part, which is not important in numerical analysis. We use the
wave functions of $\pi$ meson ( $\phi_\pi^A(x)$, $ \phi_{\pi}^P(x) $ and  $\phi_{\pi}^T(x)$ )
as given in Ref.\cite{ball3}.
For $\eta$ meson's wave function, $\phi_{\eta_{d\bar{d}}}^A$,
$\phi_{\eta_{d\bar{d}}}^P$ and $\phi_{\eta_{d\bar{d}}}^T$
represent the axial vector, pseudoscalar and tensor components of
the wave function respectively, for which we utilize the result
from the light-cone sum rule \cite{ball} including twist-3
contribution. For the explicit expressions of the wave functions and the values of related
quantities, one can see Eqs.(50) and (51) of Ref.\cite{liu05}.

We assume that the wave function of $u\bar{u}$ is same as the wave function of $d\bar{d}$.
For the wave function of the $s\bar{s}$ components, we also use
the same form as $d\bar{d}$ but with $m^{s\bar{s}}_0$ and $f_y$
instead of $m^{d\bar{d}}_0$ and $f_x$, respectively. For $f_x$ and
$f_y$, we use the values as given in Ref.\cite{kf} where isospin
symmetry is assumed for $f_x$ and $SU(3)$ breaking effect is
included for $f_y$:
 \beq
 f_x=f_{\pi}, \ \ \ f_y=\sqrt{2f_K^2-f_{\pi}^2}.\ \ \
\label{eq:7-5}
\eeq
These values are translated to the values in the two mixing angle
method, which is often used in vacuum saturation approach as:
\beq
f_8 &=&169  {\rm MeV}, \quad f_1=151  {\rm MeV},  \non
\theta_8&=& -25.9^{\circ} (-18.9^{\circ}), \quad \theta_1=-7.1^{\circ}
(-0.1^{\circ}),
\eeq
where the pseudoscalar mixing angle $\theta_p$ is taken as
$-17^{\circ}$ ($-10^{\circ}$) \cite{ekou01}.
The parameters $m_0^i$
$(i=\eta_{d\bar{d}(u\bar{u})}, \eta_{s\bar{s}})$ are defined as:
\beq
m_0^{\eta_{d\bar{d}(u\bar{u})}}\equiv m_0^\pi \equiv
\frac{m_{\pi}^2}{(m_u+m_d)}, \qquad m_0^{\eta_{s\bar{s}}}\equiv
\frac{2M_K^2-m_{\pi}^2}{(2m_s)}.
 \label{eq:19}
\eeq

We include full expression of twist$-3$ wave functions for light
mesons. The twist$-3$ wave functions are also adopted from QCD sum
rule  calculations \cite{bf}. We will see later that this set of
parameters will give good results for $B_s \to \pi^0
\eta^{(\prime)}$ decays.

\subsection{Branching ratios}

For $B_s \to \pi^0 \etap$ decays, the decay amplitudes in
Eqs.~(\ref{eq:m2}) can be rewritten as
\beq
{\cal M} &=& V_{ub}^*V_{us} T -V_{tb}^* V_{ts} P= V_{ub}^*V_{us} T
\left [ 1 + z e^{ i ( \gamma + \delta ) } \right], \label{eq:ma}
\eeq where \beq z=\left|\frac{V_{tb}^* V_{ts}}{ V_{ub}^*V_{us} }
\right| \left|\frac{P}{T}\right| \label{eq:zz}
\eeq
is the ratio of penguin to tree contributions, $\gamma = \arg
\left[-\frac{V_{ts}V_{tb}^*}{V_{us}V_{ub}^*}\right]$ is the weak
phase (one of the three CKM angles), and $\delta$ is the relative
strong phase between penguin (P) and tree (T) diagrams.
In the pQCD approach, it is easy to calculate the ratio $z$ and the
strong phase $\delta$ for the decay in study. For  $B_s \to \pi^0
\eta$ and $\pi^0 \eta'$ decays, we find numerically that
\beq
z(\pi^0\eta) &=&38.3, \qquad \delta (\pi^0\eta)=-94^\circ , \label{eq:zd1}\\
z(\pi^0\eta^\prime) &=&5.5, \qquad
\delta(\pi^0\eta^\prime)=-20^\circ .\label{eq:zd2}
\eeq
The main error of the ratio $z$ and the strong phase $\delta$ is induced
by the uncertainty of $\omega_{b_s}=0.50 \pm 0.05$ GeV.
Since the errors induced by the uncertainties of most input parameters
are largely canceled in the ratio, we will use the central
values of $z$ and $\delta$ in the following numerical
calculations, unless explicitly stated otherwise.

Using  the wave functions and the input parameters as specified in
previous sections,  it is straightforward  to calculate the
branching ratios for the four considered decays. The theoretical
predictions in the pQCD approach for the CP-averaged branching ratios of the
decays under consideration are the following
\beq
Br(\ B_s^0 \to \pi^0 \eta) &=& \left [ 0.86 ^{+0.37}_{-0.24}(\omega_{b_s}) ^{+0.33}_{-0.21}(m_s)
^{+1.00}_{-0.09} (a_t) \right ]\times 10^{-7},
\label{eq:br0-eta} \\
Br(\ B_s^0 \to \pi^0 \eta^{\prime}) &=& \left [ 1.86^{+0.76}_{-0.51} (\omega_{b_s})
^{+0.63}_{-0.41} (m_s)^{+1.46}_{-0.21} (a_t) \right ]\times 10^{-7}
\label{eq:br0-etap} ,
\eeq
for $\theta_p=-17^\circ$, and
\beq
Br(\ B_s^0 \to \pi^0 \eta) &=&\left [ 1.18^{+0.50}_{-0.33} (\omega_{b_s})
^{+0.45}_{-0.29} (m_s)^{+1.03}_{-0.12} (a_t)\right ]\times 10^{-7} ,
\label{eq:br0-eta1}\\
 Br(\ B_s^0 \to \pi^0 \eta^{\prime}) &=&\left [ 1.54
^{+0.63}_{-0.42} (\omega_{b_s}) ^{+0.52}_{-0.34} (m_s)^{+1.19}_{-0.21} (a_t)\right ] \times
10^{-7}. \label{eq:br0-etap1}
\eeq
for $\theta_p=-10^\circ$. The
main errors are induced by the uncertainties of  $a_t=1.0 \pm 0.2$, $\omega_{b_s}=0.50
\pm 0.05$ GeV and $m_s = 120 \pm 20$ MeV, respectively.

It is easy to see that (a) the errors of the branching ratios
induced by varying $a_t$ in the range of $a_t=[0.8, 1.2]$  can be
significant for the penguin-dominated $B_s \to \pi^0 \etap$
decays; and (b) the variations with respect to the central values
are large for the case of $a_t=0.8$, but very small for the case of
$a_t=1.2$). This feature agrees with general expectations: when the scale $t$
become smaller, the reliability of the perturbative calculation of
the form factors in pQCD approach will become weak!

The pQCD predictions of the branching ratios as given in Eqs.(\ref{eq:br0-eta}-\ref{eq:br0-etap1})
agree  well with the theoretical predictions in the QCDF approach,
for example, as given in Ref.~\cite{bn03b}:
\beq
Br(\ B_s^0 \to \pi^0 \eta) &=& \left ( 0.75 ^{+0.35}_{-0.30}\right ) \times
10^{-7}, \non Br(\ B_s^0 \to \pi^0 \eta^{\prime}) &=& \left ( 1.1
^{+0.24}_{-0.24}\right ) \times 10^{-7}, \label{eq:br24}
\eeq
where the individual errors as given in Ref.~\cite{bn03b} have been added in quadrature.

\subsection{CP-violating asymmetries }

Now we turn to study the CP-violating asymmetries for $B_s^0 \to
\pi^0 \etap$ decays. For these neutral decay modes, the effects of
$B_s^0-\bar{B_s}^0$ mixing should be considered.

For $B_s^0$ meson decays, we know that $\Delta \Gamma/\Delta m_s
\ll 1$ and $\Delta \Gamma/\Gamma \ll 1$. The CP-violating
asymmetry of $B_s^0(\bar B_s^0) \to \pi^0 \eta^{(\prime)}$ decay
is time dependent and can be defined as
\beq
A_{CP} &\equiv& \frac{\Gamma\left (\overline{B_s^0}(\Delta t) \to f_{CP}\right) -
\Gamma\left(B_s^0(\Delta t) \to f_{CP}\right )}{ \Gamma\left
(\overline{B_s^0}(\Delta t) \to f_{CP}\right ) + \Gamma\left
(B_s^0(\Delta t) \to f_{CP}\right ) }\non &=& A_{CP}^{dir} \cos
(\Delta m_s  \Delta t) + A_{CP}^{mix} \sin (\Delta m_s  \Delta t),
\label{eq:acp-def}
\eeq
where $\Delta m_s$ is the mass difference
between the two $B_s^0$ mass eigenstates, $\Delta t
=t_{CP}-t_{tag} $ is the time difference between the tagged
$B_s^0$ ($\overline{B_s}^0$) and the accompanying
$\overline{B_s}^0$ ($B_s^0$) with opposite b flavor decaying to
the final CP-eigenstate $f_{CP}$ at the time $t_{CP}$. The direct
and mixing induced CP-violating asymmetries $A_{CP}^{dir}$ and
$A_{CP}^{mix}$ can be written as
\beq
\acp^{dir}=\frac{ \left |
\lambda_{CP}\right |^2 -1 } {1+|\lambda_{CP}|^2}, \qquad
A_{CP}^{mix}=\frac{ 2Im (\lambda_{CP})}{1+|\lambda_{CP}|^2},
\label{eq:acp-dm}
\eeq
where the CP-violating parameter $\lambda_{CP}$ is
\beq
\lambda_{CP} = \frac{ V_{tb}^*V_{ts}
\langle \pi^0 \etap |H_{eff}| \overline{B_s}^0\rangle} {
V_{tb}V_{ts}^* \langle \pi^0 \etap |H_{eff}| B_s^0\rangle} =
e^{2i\gamma}\frac{ 1+z e^{i(\delta-\gamma)} }{
1+ze^{i(\delta+\gamma)} }. \label{eq:lambda2}
\eeq
Here the ratio $z$ and the strong phase $\delta$ have been defined previously. In
pQCD approach, since both $z$ and $\delta$ are calculable, it is
easy to find the numerical values of $A_{CP}^{dir}$ and
$A_{CP}^{mix}$ for the considered decay processes.

\begin{figure}[htb]
\centerline{\mbox{\epsfxsize=10cm\epsffile{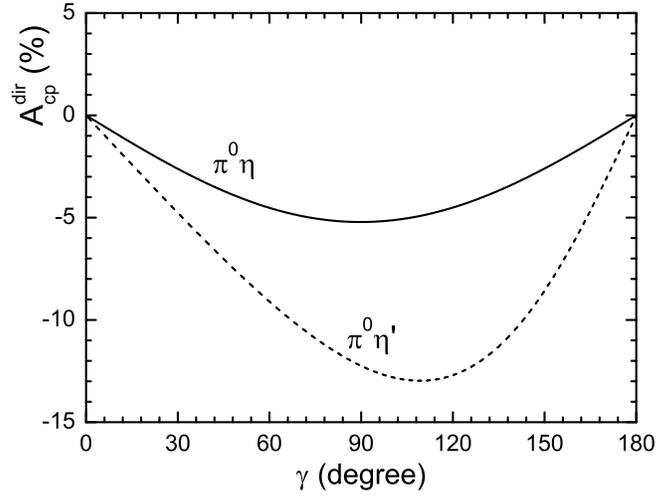}}}
\vspace{0.2cm}
\caption{The direct CP asymmetry $A_{CP}^{dir}$ (in percentage) of
 $B_s\to \pi^0 \eta$ (solid curve)
 and $B_s\to \pi^0 \eta^{\prime}$ (dotted curve) as a function of CKM
angle $\gamma$ for the case of $\theta_p=-17^\circ$.}
\label{fig:fig2}
\end{figure}

In Figs.~\ref{fig:fig2}, we show the $\gamma-$dependence of the
direct CP-violating asymmetry $A_{CP}^{dir}$ for $B_s^0 \to \pi^0
\eta$ (solid curve) and $B_s^0 \to \pi^0 \eta^\prime$ (dotted
curve) decays for $\theta_p=-17^\circ$ .

\begin{figure}[htb]
\centerline{\mbox{\epsfxsize=10cm\epsffile{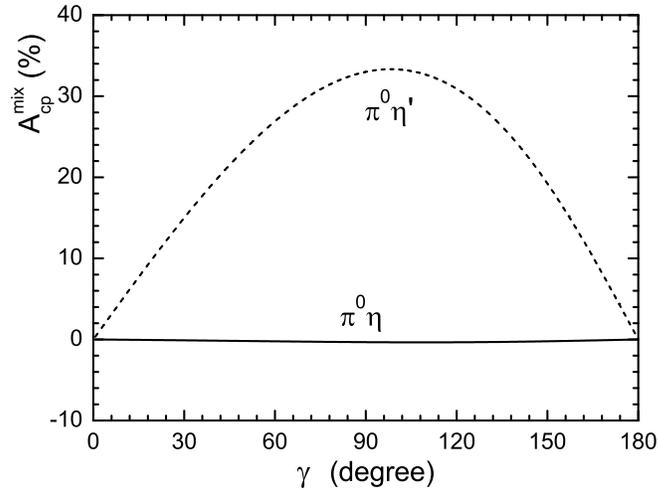}}}
\vspace{0.2cm} \caption{The mixing induced CP asymmetry
$A_{CP}^{mix}$ (in percentage) of $B_s\to \pi^0 \eta$ (solid
curve) and $B_s\to \pi^0 \eta^{\prime}$ (dotted curve) as a
function of CKM angle $\gamma$ for the case of $\theta_p=-17^\circ$ .}
\label{fig:fig3}
\end{figure}

The pQCD predictions for the direct CP-violating asymmetries of
$B_s^0 \to \pi^0 \etap$ decays are
\beq
\acp^{dir}(B_s^0 \to \pi^0 \eta) &=& \left [ -4.5 ^{+1.2}_{-0.6}(\gamma) ^{+0.6}_{-0.4}(\omega_{b_s})
\pm 0.6 (m_0^\pi)^{+1.7}_{-1.8} (m_s)^{+0.7}_{-0.2}(a_t)
  \right ] \times 10^{-2} \label{eq:acp-d1}, \\
\acp^{dir}(B_s^0 \to \pi^0 \eta^\prime) &=& \left [ -9.1 ^{+2.8}_{-2.3}(\gamma)
^{+0.3}_{-0.6}(\omega_{b_s})
\pm 0.3 (m_0^\pi) \pm 1.9 (m_s)^{+4.1}_{-1.5} (a_t) \right ]
\times 10^{-2}
\label{eq:acp-d2}.
\eeq

As a comparison, we present the QCDF predictions for
$\acp^{dir}(B_s^0 \to \pi^0 \eta')$ directly quoted from
Ref.~\cite{bn03b}
\beq
\acp^{dir}(B_s^0 \to \pi^0 \eta^\prime) &=&
\left (27.8^{+6.0\; +9.6\; +2.0\; +24.7}_{ -7.1\; -5.7\; -2.0 \;
-27.2} \right ) \times 10^{-2}\label{eq:acp-db},
\eeq
where the ``default values" of the input parameters have been used
in Ref.~\cite{bn03b}, and the error sources are the same as the first four
input parameters in Eqs.~(\ref{eq:acp-d1}) and (\ref{eq:acp-d2}).
Currently, no relevant experimental measurements for the
CP-violating asymmetries of $B_s^0 \to \pi^0 \etap$ decays are
available. For the direct CP-violating asymmetries of $B_s^0 \to
\pi^0 \etap$ decays, the theoretical predictions in pQCD and QCDF
approach have the opposite sign, but the theoretical errors are
clearly too large to make a meaningful comparison. One has to wait
for the improvements in both the experimental measurements and the
calculation of high order contributions.

The pQCD predictions for the mixing induced CP-violating
asymmetries of $B_s^0 \to \pi^0 \etap$ decays  are
\beq
\acp^{mix}(B_s^0 \to \pi^0 \eta) &=& \left [ -0.2
\pm 0.1 (\gamma) ^{+2.5}_{-2.1}(\omega_{b_s})
^{+1.2}_{-1.4}(m_0^\pi)^{+4.4}_{-4.5}(m_s)^{+26.3}_{-11.6} (a_t)
\right ] \times 10^{-2} \label{eq:acp-m1}, \\
\acp^{mix}(B_s^0 \to \pi^0 \eta^\prime) &=& \left [ 27.0
^{+4.8}_{-7.5}(\gamma) ^{+0.4}_{-0.7} (\omega_{b_s})
^{+0.6}_{-0.5}(m_0^\pi) \pm 0.2  (m_s)^{+17.1}_{-8.3} (a_t)
\right ] \times 10^{-2} \label{eq:acp-m2},
\eeq
where the dominant errors come from the variations of $\omega_{b_s}=0.50\pm 0.05$
GeV, $m_0^\pi=1.4\pm 0.3$ GeV, $a_t=1.0 \pm 0.2$, $m_s = 120 \pm
20$MeV and $\gamma=60^\circ \pm 20^\circ$.

If we integrate the time variable $t$, we will get the total CP
asymmetry for $B_s^0 \to \pi^0 \etap$ decays,
\beq
A_{CP}=\frac{1}{1+x^2} A_{CP}^{dir} + \frac{x}{1+x^2}
A_{CP}^{mix},
\eeq
where $x=\Delta m_s/\Gamma=26.5$ for the $B_s^0-\overline{B_s}^0$ mixing \cite{0603003}.
We found  numerically that the magnitude of the total CP asymmetry for
$ (B_s^0 \to \pi^0 \etap)$ decays are smaller than $2\%$ in the whole considered parameter space.

\subsection{Effects of possible gluonic component of $\eta^\prime$}

Up to now, we have not considered the possible  contributions to
the branching ratios and CP-violating asymmetries of $B_s^0 \to
\pi \eta^\prime$ decays induced by the possible gluonic component
of $\eta^\prime$ \cite{rosner83,ekou01}. When
$Z_{\eta^\prime} \neq 0$, a decay amplitude ${\cal M}'$ will be
produced by the gluonic component of $\eta^\prime$. Such decay
amplitude may construct or destruct with the ones from the
$q\bar{q}$ ($q=u,d,s$) components of $\eta^\prime$, the branching
ratios of the decays in question may be increased or decreased
accordingly.

Unfortunately, we currently do not know how to calculate this kind
of contributions reliably. But we can treat it as an theoretical
uncertainty. For $|M'/M(q\bar{q})| \sim 0.1-0.2$, for example, the
resulted uncertainty for the branching ratios as given in
Eq.(\ref{eq:br0-etap}) will be around twenty to thirty percent.

Furthermore, the pQCD predictions for the branching ratios of $B
\to \rho \etap$ and $B \to \pi \etap$ decays also show very good
agreement with the data \cite{liu05,wang05}. We therefore believe
that the gluonic admixture of $\eta^\prime$ should be small, and
most possibly not as important as expected before.

As for the CP-violating asymmetries of $B_s^0 \to \pi^0
\eta^\prime$ decays, the possible contributions of the gluonic
components of the $\eta^\prime$ meson are largely canceled in the
ratio. These results may be measured in the forthcoming LHCb
experiments.

\section{summary }

In this paper,  we calculate the branching ratios and CP-violating
asymmetries of $B_s^0 \to \pi^0 \eta$, $B_s^0 \to \pi^0
\eta^{\prime}$ decays in the pQCD factorization approach.

Besides the usual factorizable diagrams, the non-factorizable and
annihilation diagrams are also calculated analytically. Although
the non-factorizable and annihilation contributions are
sub-leading for the branching ratios of the considered decays, but
they are not negligible. Furthermore these diagrams provide the
necessary strong phase required by a non-zero CP-violating
asymmetry for the considered decays.

From our calculations and phenomenological analysis, we found the following
results:
\begin{itemize}

\item
The pQCD predictions for the form factors are $F_{0,1}^{B_s \to
\eta}(0)=-0.276$ and $F_{0,1}^{B_s \to \eta^\prime}(0)=0.278$, which agree well
with those obtained from other methods.

\item
For the CP-averaged branching ratios of the considered
decay modes, the pQCD predictions for $\theta_p=17^\circ$ are
\beq
Br(B_s^0 \to \pi^0 \eta ) &=&\left ( 0.86^{+1.12}_{-0.33} \right )
\times 10^{-7}, \non
Br(B_s^0 \to \pi^0 \eta^{\prime}) &=&\left (
1.86^{+1.76}_{-0.69} \right ) \times 10^{-7},
\eeq
here the various errors as specified in Eqs.~(\ref{eq:br0-eta}) and (\ref{eq:br0-etap})
have been added in quadrature. The pQCD predictions are also well consistent with the results
obtained by employing the QCD factorization approach.

\item
For the CP-violating asymmetries, the pQCD predictions for $\acp^{dir}(B_s\to \pi^0 \etap)$
and $\acp^{mix}(B_s\to \pi^0 \etap)$ are generally not very large, while the time-integrated CP
asymmetries are less than $2\%$ in magnitude.

\item
The major theoretical errors of the computed observables are
induced by the uncertainties of the hard energy scale $t_j$'s, the
parameters $\omega_{b_s}$ and $m_s$, as well as the CKM angle $\gamma$ for CP asymmetries.
\end{itemize}

\begin{acknowledgments}

This work is partly supported  by the National Natural Science
Foundation of China under Grant No.10275035, 10575052, and by the
Specialized Research Fund for the doctoral Program of higher education (SRFDP)
under Grant No.~20050319008.

\end{acknowledgments}


\begin{appendix}

\section{Related Functions }\label{sec:aa}

We show here the function $h_i$'s, coming from the Fourier
transformations  of $H^{(0)}$,
\beq
 h_e(x_1,x_3,b_1,b_3)&=&
 K_{0}\left(\sqrt{x_1 x_3} m_B b_1\right)
 \left[\theta(b_1-b_3)K_0\left(\sqrt{x_3} m_B
b_1\right)I_0\left(\sqrt{x_3} m_B b_3\right)\right.
 \non
& &\;\left. +\theta(b_3-b_1)K_0\left(\sqrt{x_3}  m_B b_3\right)
I_0\left(\sqrt{x_3}  m_B b_1\right)\right] S_t(x_3), \label{he1}
\eeq
 \beq
 h_a(x_2,x_3,b_2,b_3)&=&
 K_{0}\left(i \sqrt{x_2 x_3} m_B b_3\right)
 \left[\theta(b_3-b_2)K_0\left(i \sqrt{x_3} m_B
b_3\right)I_0\left(i \sqrt{x_3} m_B b_2\right)\right.
 \non
& &\;\;\;\;\left. +\theta(b_2-b_3)K_0\left(i \sqrt{x_3}  m_B
b_2\right) I_0\left(i \sqrt{x_3}  m_B b_3\right)\right] S_t(x_3),
\label{he3} \eeq
 \beq
 h_{f}(x_1,x_2,x_3,b_1,b_2) &=&
 \biggl\{\theta(b_2-b_1) \mathrm{I}_0(M_B\sqrt{x_1 x_3} b_1)
 \mathrm{K}_0(M_B\sqrt{x_1 x_3} b_2)
 \non
&+ & (b_1 \leftrightarrow b_2) \biggr\}  \cdot\left(
\begin{matrix}
 \mathrm{K}_0(M_B F_{(1)} b_1), & \text{for}\quad F^2_{(1)}>0 \\
 \frac{\pi i}{2} \mathrm{H}_0^{(1)}(M_B\sqrt{|F^2_{(1)}|}\ b_1), &
 \text{for}\quad F^2_{(1)}<0
\end{matrix}\right),
\label{eq:pp1}
 \eeq
\beq
h_f^3(x_1,x_2,x_3,b_1,b_2) &=& \biggl\{\theta(b_1-b_2)
\mathrm{K}_0(i \sqrt{x_2 x_3} b_1 M_B)
 \mathrm{I}_0(i \sqrt{x_2 x_3} b_2 M_B)+(b_1 \leftrightarrow b_2) \biggr\}
 \non
& & \cdot
 \frac{\pi i}{2} \mathrm{H}_0^{(1)}(\sqrt{x_1+x_2+x_3-x_1 x_3-x_2 x_3}\ b_1 M_B),
 \label{eq:pp4}
\eeq
 \beq
 h_f^4(x_1,x_2,x_3,b_1,b_2) &=&
 \biggl\{\theta(b_1-b_2) \mathrm{K}_0(i \sqrt{x_2 x_3} b_1 M_B)
 \mathrm{I}_0(i \sqrt{x_2 x_3} b_2 M_B)
 \non
&+& (b_1 \leftrightarrow b_2) \biggr\} \cdot \left(
\begin{matrix}
 \mathrm{K}_0(M_B F_{(2)} b_1), & \text{for}\quad F^2_{(2)}>0 \\
 \frac{\pi i}{2} \mathrm{H}_0^{(1)}(M_B\sqrt{|F^2_{(2)}|}\ b_1), &
 \text{for}\quad F^2_{(2)}<0
\end{matrix}\right), \label{eq:pp3}
\eeq
where $J_0$ is the Bessel function and  $K_0$, $I_0$ are
modified Bessel functions $K_0 (-i x) = -(\pi/2) Y_0 (x) + i
(\pi/2) J_0 (x)$, and $F_{(j)}$'s are defined by
\beq
F^2_{(1)}&=&(x_1 -x_2) x_3\;,\\
F^2_{(2)}&=&(x_1-x_2) x_3\;\;.
 \eeq

The threshold resummation form factor $S_t(x_i)$ is adopted from
Ref.\cite{kurimoto}
\beq S_t(x)=\frac{2^{1+2c} \Gamma
(3/2+c)}{\sqrt{\pi} \Gamma(1+c)}[x(1-x)]^c,
\eeq
where the parameter $c=0.3$. This function is normalized to unity.

The Sudakov factors used in the text are defined as \beq S_{ab}(t)
&=& s\left(x_1 m_B/\sqrt{2}, b_1\right) +s\left(x_3 m_B/\sqrt{2},
b_3\right) +s\left((1-x_3) m_B/\sqrt{2}, b_3\right) \non
&&-\frac{1}{\beta_1}\left[\ln\frac{\ln(t/\Lambda)}{-\ln(b_1\Lambda)}
+\ln\frac{\ln(t/\Lambda)}{-\ln(b_3\Lambda)}\right],
\label{wp}\\
S_{cd}(t) &=& s\left(x_1 m_B/\sqrt{2}, b_1\right)
 +s\left(x_2 m_B/\sqrt{2}, b_2\right)
+s\left((1-x_2) m_B/\sqrt{2}, b_2\right) \non
 && +s\left(x_3
m_B/\sqrt{2}, b_2\right) +s\left((1-x_3) m_B/\sqrt{2}, b_2\right)
\non
 & &-\frac{1}{\beta_1}\left[2
\ln\frac{\ln(t/\Lambda)}{-\ln(b_1\Lambda)}
+\ln\frac{\ln(t/\Lambda)}{-\ln(b_2\Lambda)}\right],
\label{Sc}\\
S_{ef}(t) &=& s\left(x_1 m_B/\sqrt{2}, b_1\right)
 +s\left(x_2 m_B/\sqrt{2}, b_2\right)
+s\left((1-x_2) m_B/\sqrt{2}, b_2\right) \non
 && +s\left(x_3
m_B/\sqrt{2}, b_2\right) +s\left((1-x_3) m_B/\sqrt{2}, b_2\right)
\non
 &
&-\frac{1}{\beta_1}\left[\ln\frac{\ln(t/\Lambda)}{-\ln(b_1\Lambda)}
+2\ln\frac{\ln(t/\Lambda)}{-\ln(b_2\Lambda)}\right],
\label{Se}\\
S_{gh}(t) &=& s\left(x_2 m_B/\sqrt{2}, b_2\right)
 +s\left(x_3 m_B/\sqrt{2}, b_3\right)
+s\left((1-x_2) m_B/\sqrt{2}, b_2\right) \non
 &+& s\left((1-x_3)
m_B/\sqrt{2}, b_3\right)
-\frac{1}{\beta_1}\left[\ln\frac{\ln(t/\Lambda)}{-\ln(b_1\Lambda)}
+\ln\frac{\ln(t/\Lambda)}{-\ln(b_2\Lambda)}\right], \label{ww}
\eeq where the function $s(q,b)$ are defined in the Appendix A of
Ref.\cite{luy01}.

\end{appendix}


\end{document}